\documentstyle[aps,epsf]{revtex}

\newcommand\g{\gamma}
\renewcommand\d{\delta}
\newcommand\e{\epsilon}

\newcommand\q{\theta}

\renewcommand\k{\kappa}

\newcommand\m{\mu}
\newcommand\n{\nu}

\newcommand\p{\pi}

\newcommand\f{\phi}
\newcommand\vf{\varphi}

\renewcommand\o{\omega}

\newcommand\D{\Delta}

\newcommand\G{\Gamma}
\newcommand\F{\Phi}

\newcommand\ra{\rightarrow}

\newcommand{\be}{\begin{equation}}
\newcommand{\ee}{\end{equation}}
\newcommand{\bea}{\begin{eqnarray}}
\newcommand{\eea}{\end{eqnarray}}
\newcommand{\ba}[1]{\begin{array}{#1}}
\newcommand{\ea}{\end{array}}
\newcommand{\non}{\nonumber \\}

\newcommand{\eqrf}[1]{Eq.\ (\ref{#1})}

\newcommand{\dirac}[2]{#1\llap{/\kern#2}}

\begin{document}

\title{How the quark self-energy affects the 
color-superconducting gap}

\author{Qun Wang}
\address
{
\small\it Institut f\"ur Theoretische Physik,
J.W.\ Goethe-Universit\"at, D-60054 Frankfurt am Main, Germany\\
and Physics Department, Shandong University,
Jinan, Shandong 250100, P. R. China \\
}

\author{Dirk H. Rischke}

\address 
{
\small\it Institut f\"ur Theoretische Physik,
J.W.\ Goethe-Universit\"at, D-60054 
Frankfurt am Main, Germany
}

\date{\today} 

\maketitle

\begin{abstract}

We consider color superconductivity with 
two flavors of massless quarks which form 
Cooper pairs with total spin zero. 
We solve the gap equation for 
the color-superconducting gap parameter 
to subleading order in the QCD coupling constant 
$g$ at zero temperature. At this order in $g$, 
there is also a previously neglected 
contribution from the real part of the 
quark self-energy to the gap equation. 
Including this contribution leads to 
a reduction of the color-superconducting gap 
parameter $\f_0$ by a factor 
$b_0'=\exp \big[ -(\p ^2+4)/8 \big]\simeq 0.177$. 
On the other hand, the BCS relation 
$T_c\simeq 0.57\f_0$ between $\f_0$ and 
the transition temperature $T_c$ 
is shown to remain valid after 
taking into account corrections 
from the quark self-energy. 
The resulting value for $T_c$ 
confirms a result obtained 
previously with a different method.

\end{abstract}

\section{Introduction}

Quantum chromodynamics (QCD) is the fundamental theory of the strong
interaction. In strongly interacting matter at large density or, 
equivalently, large quark chemical potential $\m$, 
asymptotic freedom \cite{asymp} implies that 
single-gluon exchange becomes the dominant interaction between quarks.
Single-gluon exchange is attractive in the color-antitriplet 
channel \cite{bai84}. By Cooper's theorem \cite{BCS}, 
any attractive interaction destabilizes the Fermi surface and, 
at sufficiently small temperature $T$, leads to the condensation of 
Cooper pairs. If the Cooper pair condensate carries charge quantum numbers
of a local gauge symmetry, the Meissner effect leads to superconductivity. 
Strongly interacting matter, where quark Cooper pairs carry color
charge, becomes a color superconductor. In a superconductor, exciting
particle-hole pairs costs at least an energy amount 
$2\f_0$, where $\f_0$ is the value of the superconducting
gap parameter at the Fermi surface for $T=0$.
Its value can be computed from a gap equation 
derived in mean-field approximation, which, in QCD, involves 
single-gluon exchange \cite{SchaferWilczek,pis00}.

Schematically, this gap equation can be written 
in the form \cite{ris01}
\be
\f_0 = g^2 \, \left[\; \zeta \,
\ln^2 \left( \frac{\m}{\f_0} \right) 
+  \beta \, \ln \left( \frac{\m}{\f_0} \right) +  \alpha\;
\right] \; \f_0\;.
\label{gapequation}
\ee
For small values of the QCD coupling constant, 
$g \ll 1$, the solution is 
\cite{SchaferWilczek,pis00,son99,hong,miransky}
\be
\f_0 = 2 \, b \, \m \, \exp \left( - \frac{c}{g} \right) 
\left[ 1 + O(g) \right]\;.
\label{gapsolution}
\ee
The first term in \eqrf{gapequation} contains two powers of the logarithm 
$\ln (\m/\f_0)$. One logarithm is well-known from the gap equation
in standard BCS theory \cite{BCS}, 
where it arises from the integration over 
fermion momenta up to the Fermi surface.
The other logarithm is special to
theories with long-range interactions, like 
the exchange of almost static magnetic gluons in QCD  
\cite{pis00,son99}. Its origin is a collinear singularity when
integrating over angles between quark and gluon momenta 
in the gap equation. 
The weak coupling solution (\ref{gapsolution}) 
implies that the first term in
brackets in \eqrf{gapequation} is $\sim 1/g^2$. 
It therefore dominates
the right-hand side of \eqrf{gapequation}. Together with 
the prefactor $g^2$, it is of order $O(1)$ in the gap equation. 
The value of the coefficient $\zeta$ determines 
the constant $c$ in \eqrf{gapsolution}.

The second term in \eqrf{gapequation} contains subleading
contributions of order $O(g)$ to the gap equation, characterized by a
single power of the logarithm $\ln (\m/\f_0) \sim 1/g$.
A part of these contributions arises from the
exchange of non-static magnetic and static electric
gluons \cite{pis00}. Both types of interactions are short-range:
they are screened on a distance scale $m_g^{-1}$, where $m_g$ is the
gluon mass; $m_g^2=N_f g^2 \m ^2 /(6\p ^2)$, $N_f$ is the number of
quark flavors. Consequently, the collinear logarithm characteristic for
long-range interactions is absent, and one is left with 
the BCS logarithm. The coefficient $\beta$ in \eqrf{gapequation}
determines the constant $b$ in \eqrf{gapsolution}.

The third term in \eqrf{gapequation} summarizes sub-subleading
contributions of order $O(g^2)$ 
with neither a collinear nor a BCS logarithm.
It was argued in \cite{SchaferWilczek,pis00,ShusterRajagopal} that
at this order gauge-dependent terms enter the QCD gap equation.
However, the gap parameter is in principle an observable quantity, 
and thus gauge-independent. Therefore, one concludes that the  
mean-field approach cannot be used to 
compute sub-subleading contributions to 
the gap parameter. It was also shown \cite{man1} that
effects from the finite lifetime of quasiparticles in the Fermi
sea influence the value of $\f_0$ at this order. 
In weak coupling, these contributions are suppressed by one power
of $g$ compared to the subleading terms and therefore constitute an
order $O(g)$ correction to the prefactor $b$, as indicated in
\eqrf{gapsolution}.

The value of the coefficient $c$ was first computed by Son
\cite{son99}, 
\be
\frac{c}{g} = \frac{\p}{2 \bar{g}}\;, \; \; 
\bar{g} \equiv \frac{g}{3 \sqrt{2}} \;.
\label{c}
\ee
Son also gave an estimate for the constant $b$, 
\be
b = \frac{b_0}{g^5}\;,
\label{b}
\ee 
with a constant $b_0$ of order $O(1)$, which could not be determined in the
approach of Ref.\ \cite{son99}.
In \cite{SchaferWilczek,pis00} the constant $b_0$ was
computed by solving the QCD gap equation including 
non-static magnetic and static electric gluon exchange. The result is
\be
\label{b0}
b_0 = 256\, \pi^4 \left( \frac{2}{N_f} \right)^{5/2}\ b_0'\;,
\ee
with an undetermined constant $b_0'$ of order $O(1)$.
In \cite{SchaferWilczek,pis00}, where the only subleading
contributions to the gap equation arise from non-static magnetic
and static electric gluon exchange, $b_0'=1$. In principle, however,
there could be other subleading contributions, which would
change $b_0'$ to a value $b_0' \neq 1$.

At sufficiently large temperature, thermal random motion breaks up
Cooper pairs and the superconducting condensate melts.
In Ref.\ \cite{pis00}, it was shown that the 
temperature $T_c$ for the transition between the normal and the 
superconducting phase is related to the zero-temperature gap at the
Fermi surface in the same way as in BCS theory,
\be
T_c = \frac{e^\gamma}{\p}\, \f_0 \simeq 0.57 \, \f_0\; ,
\label{Tcphi}
\ee
where $\gamma \simeq 0.577$ is the Euler-Mascheroni constant.

In Ref.\ \cite{bro00}, Brown, Liu, and Ren 
calculated $T_c$ in a different approach with the result
\be
T_c = 2\, \frac{e^\gamma}{\p} \, 256\, \pi^4 \left( \frac{2}{N_f g^2} 
\right)^{5/2}\, c_1' \, \mu \,
\exp \left( - \frac{\p}{2 \bar{g}} \right) \;,
\label{Tc}
\ee
where 
\be
c_1' = \exp \left( - \frac{\p^2 + 4}{8} \right) \simeq 0.177\; .
\label{c1'}
\ee
Furthermore, the authors of \cite{bro00} assumed 
the validity of \eqrf{Tcphi}, and concluded that
\be
b_0' = c_1'\; ,
\label{b0'}
\ee
as one readily checks with 
Eqs.\ (\ref{gapsolution})--(\ref{b0}), (\ref{Tc}), and
(\ref{c1'}). Physically,
the difference between the approach of
Refs.\ \cite{SchaferWilczek,pis00} and that of \cite{bro00} is that
contributions from the quark self-energy were
neglected in the former, but taken into account in the latter. 
If the above arguments are correct,
one may therefore conclude that the quark self-energy constitutes
a subleading correction to the gap equation and thus is responsible
for the change of the value of $b_0'$ from 1 to $c_1'$ given by 
\eqrf{c1'}. The authors of \cite{bro00} 
also assert that there are no further
subleading contributions that could alter the value of $c_1'$.

While manifestly gauge invariant, a disadvantage of the approach of 
\cite{bro00} is that its range of applicability is restricted to
the normal-conducting phase and thus can only determine
the value of $T_c$, but not the value of the zero-temperature
gap $\f_0$. Therefore, a relation between $T_c$ and $\f_0$, 
like \eqrf{Tcphi}, can in principle not be established 
within this approach. It is possible to derive such a relation 
with the help of the gap equation, as demonstrated in \cite{pis00}, 
but contributions from the quark self-energy were neglected in 
obtaining the result \eqrf{Tcphi}. 
It is conceivable that \eqrf{Tcphi} changes, once 
these contributions are taken into account. 
Consequently, the validity of \eqrf{b0'} is not obvious.

The aim of the present paper is twofold. 
On the one hand, we want to
compute the contribution of 
the quark self-energy to the 
value of the constant $b_0'$ in the zero-temperature
gap. On the other hand, we want to confirm the result (\ref{Tc}) 
for $T_c$. To this end,
it is necessary to first compute the value of the zero-temperature
gap by directly solving the gap equation including the quark
self-energy. Second, one has to prove that 
\eqrf{Tcphi} remains valid in order
to determine $T_c$, which then can be compared to the value (\ref{Tc})
obtained in \cite{bro00}.
Our paper is organized as follows.
In Section II we first clarify how 
the quark self-energy enters the gap equation.
In Section III, the resulting gap equation is solved at zero
temperature. In Section IV  we determine $T_c$. We conclude
in Section V with a summary of our results.

Our convention for the metric tensor is $g^{\m\n}=
{\rm diag}\{1,-1,-1,-1\}$. Our units are $\hbar=c=k_B=1$. 
Four-vectors are denoted by capital letters,
$K \equiv K^\mu = (k_0,{\bf k})$, 
and $k \equiv |{\bf k}|$, while 
$\hat{\bf k} \equiv {\bf k}/k$.

\section{The gap equation including the quark self-energy}

In fermionic systems at non-zero density, 
it is advantageous to treat fermions and 
charge-conjugate fermions as independent degrees of
freedom and to work in the so-called Nambu-Gorkov basis. 
In this basis, the full inverse fermion 
propagator is defined as \cite{man1}
\bea
&&S^{-1}\equiv\left(\ba{cc} S_{11}^{-1}&S_{12}^{-1}\\
                S_{21}^{-1}&S_{22}^{-1} 
        \ea \right)
=\left(\ba{cc} {S^0}_{11}^{-1} + \Sigma_{11} & \Sigma_{12}\\ 
               \Sigma_{21} & {S^0}_{22}^{-1} + \Sigma_{22} 
       \ea \right)\;,
\label{S1-2}
\eea
where ${S^0}_{11}$ is the propagator for free fermions, 
${S^0}_{22}$ the propagator for free 
charge-conjugate fermions. In momentum space 
and for $\m \gg m$,
\bea 
{S^0}_{11}(Q)=\left( \g^{\m}Q_{\m} + \m \g_0 \right)^{-1}\;,\;\;
{S^0}_{22}(Q)=\left( \g^{\m}Q_{\m} - \m \g_0 \right)^{-1}\;, 
\eea
where $\g ^{\m}$ are Dirac matrices. 
The four components of the fermion self-energy 
are denoted as $\Sigma_{ij}$, $i,j=1,2$. 
The $11$ component of the self-energy, 
$\Sigma_{11}$, is the standard one-loop 
self-energy for fermions; similarly, $\Sigma_{22}$ 
is the self-energy for charge-conjugate fermions. 
The $21$ component of the self-energy, 
$\Sigma_{21}$, which was denoted $\F^+$ in \cite{pis00}, 
is the gap matrix in a superconductor, 
while $\Sigma_{12}=\g _0\Sigma_{21}^{\dagger}\g _0$. 
In the following, we somewhat imprecisely use 
the term "self-energy" only for the  
diagonal components $\Sigma_{11}$ and $\Sigma_{22}$.  

Inverting \eqrf{S1-2} one obtains the full 
fermion propagator $S$, with the diagonal 
components 
\begin{mathletters}
\label{S}
\bea
S_{11}=\left[ {S^0}_{11}^{-1} + \Sigma_{11} 
-\Sigma_{12} ({S^0}_{22}^{-1} + \Sigma_{22})^{-1}
\Sigma_{21}\right]^{-1}\;,
\eea
describing the (normal) propagation of fermions, and 
\bea
S_{22}=\left[ {S^0}_{22}^{-1} + \Sigma_{22} 
-\Sigma_{21} ({S^0}_{11}^{-1} + \Sigma_{11})^{-1}
\Sigma_{12}\right]^{-1}\;,  
\eea
describing the (normal) propagation 
of charge-conjugate fermions. 
In superconductors, due to the presence 
of a fermion-fermion condensate one can always
convert an incoming fermion into an outgoing 
charge-conjugate fermion and vice versa. Therefore,  
the full fermion propagator $S$ 
also has off-diagonal components,  
\bea
S_{12}& = & - ({S^0}_{11}^{-1} + \Sigma_{11})^{-1}\Sigma_{12}S_{22}\;,\\
S_{21}& = & - ({S^0}_{22}^{-1} + \Sigma_{22})^{-1}\Sigma_{21}S_{11}\;,  
\eea
\end{mathletters}
describing the {\em anomalous\/} propagation of fermions and 
charge-conjugate fermions.
[Please note that our sign convention 
for the self-energy differs from that in \cite{man1}, 
which leads to the difference   
between our Eqs.\ (\ref{S}) 
and Eqs.\ (2.4) and (2.5) in 
\cite{man1}.]

Let us now consider a system of quarks 
interacting via one-gluon exchange. 
In mean-field approximation \cite{pis99-2}, 
the four components of the fermion self-energy 
in momentum space are computed as
\bea
\Sigma_{ij}(K)=-g^2\frac{T}{V}\sum_Q
\D_{\m\n}^{ab}(K-Q)\left[ \hat{\G}_a^{\m} 
S(Q)\hat{\G}_b^{\n} \right]_{ij}\;,\;\;i,j=1,2\;. 
\label{Sig}
\eea
Here, $\D_{\m\n}^{ab}$ is the gluon propagator, 
and $\hat{\G}_a^{\m}$ is 
the diagonal Nambu-Gorkov matrix
$\hat{\G}_a^{\m}={\rm diag} 
(\g ^{\m}T_a, -\g ^{\m}T_a^T)$, 
$T_a$ are the Gell-Mann matrices. 
We compute the self-energy 
in the imaginary-time formalism, 
i.e., $T/V\sum_Q\equiv T\sum_n\int d^3{\bf q}/(2\p)^3$, 
where $n$ labels the fermionic 
Matsubara frequencies, 
$\o _n=(2n+1)\p T\equiv iq_0$.  
For $ij=11$, \eqrf{Sig} becomes 
Eq.\ (2.7) of \cite{man1},   
for $ij=21$, we recover Eq.\ (2.6) of \cite{man1}  
(however, due to our different sign 
convention, only up to an overall sign).

A fully self-consistent treatment of  
the mean-field approximation requires 
to solve the coupled system of Eqs.\ (\ref{S}) 
and (\ref{Sig}). The mean-field solution 
obtained in this way 
resums terms of infinite order 
in the coupling constant. 
However, because only a particular 
class of diagrams is taken into account 
(the so-called "rainbow" diagrams), 
such a solution is in general not gauge 
invariant. On the other hand, the quasi-particle 
properties encoded in the 
propagator, like their excitation spectrum,  
are physical observables and thus in principle 
gauge invariant. Indeed, a complete solution 
of the Schwinger-Dyson equations, 
as well as a perturbative expansion 
in powers of $g$, preserve gauge invariance.
Nevertheless, as was discussed in the introduction,
an expansion of the mean-field
equation (\ref{Sig}) for the color-superconducting
gap matrix $\Phi^+ \equiv \Sigma_{21}$ in powers of $g$ 
is believed to be 
gauge invariant up to terms of subleading order [the first
two terms in \eqrf{gapequation}], and the gauge
dependence only surfaces at sub-subleading order [the third term
in \eqrf{gapequation}, or the terms 
$\sim O(g)$ in \eqrf{gapsolution}].
To preserve gauge invariance beyond subleading order,
other diagrams than those of rainbow topology have to
be added to \eqrf{Sig}, or in other words, one has to go beyond
the mean-field approximation to solve for the gap matrix $\Sigma_{21}$.

If one restricts the computation of the gap to
subleading accuracy, however, the mean-field equation (\ref{Sig})
should be sufficient to obtain a gauge-invariant result.
It is then mandatory to identify {\em all\/} terms that
can contribute to subleading order. 
In \eqrf{Sig} for $\Sigma_{21}$, 
the term in square brackets becomes 
$-\g^{\m} T_a^T S_{21}(Q) \g^{\n} T_b$. 
With the exception of Ref.\ \cite{man1}, 
previous calculations of the QCD gap parameter 
neglected the terms 
$\Sigma_{11}$ and $\Sigma_{22}$ in $S_{21}$, 
see \eqrf{S}. A perturbative calculation 
of these self-energies, i.e., 
approximating $\left[ \hat{\G}_a^{\m} 
S(Q)\hat{\G}_b^{\n} \right]_{11}\simeq 
\g^{\m} T_a {S^0}_{11}(Q) \g^{\n} T_b$, 
and analogously for $ij=22$, and analytical 
continuation to real energies $q_0$ 
gives the result \cite{man0} 
\bea
\label{self-energy}
{\Sigma^0}(Q) \equiv 
{\Sigma^0}_{11}(Q) = {\Sigma^0}_{22}(Q)
\simeq \g_0\; \bar{g}^2 \left( 
q_0 \ln \frac{M^2}{q_0^2} 
+i\p|q_0|\right)\;, 
\eea
where $M^2=(3\p /4)m_g^2$. 
On the quasi-particle mass shell, $q_0=\e _q$, 
and near the Fermi surface, $\e_q \simeq \f_0$, 
the real part of the self-energy is of order 
$g^2 \f_0 \ln (\m /\f_0) \sim g \f_0 $, while 
the imaginary part is $\sim g^2 \f_0 $ and thus down 
by a factor of $g$ compared to the real part.

In \cite{man1}, the real part 
of ${\Sigma^0}$ was neglected 
and the effect of the imaginary part on the magnitude 
of the color-superconducting gap was studied. 
It was found that a non-vanishing imaginary part leads to 
sub-subleading corrections [terms included in the third
term $\sim \alpha$ in \eqrf{gapequation}] and to 
corrections of order $O(g)$ 
to the prefactor of the gap, cf.\ \eqrf{gapsolution}. 
Therefore, they are of the same order 
as terms that violate gauge invariance in the 
mean-field approximation 
\cite{SchaferWilczek,pis00,ShusterRajagopal}.

Since the real part of the self-energy 
is parametrically larger than the imaginary part by one power 
of $g$, we expect the former to contribute to subleading order, 
$O(g)$, to the gap equation, and therefore lead to 
a correction of order $O(1)$ to the prefactor.
As discussed in the introduction, this is precisely what
the authors of \cite{bro00} found, assuming the validity
of \eqrf{Tcphi}.
In the next section, we solve the gap equation
including the quark self-energy and compute the value
of $b_0'$ at zero temperature. In Section IV we then
check the validity of \eqrf{Tcphi}.

First note that, since the real part of the 
quark self-energy is expected to influence the value of $\f_0$ 
only at subleading order in the gap equation, 
it is sufficient to approximate the value 
of $\Sigma_{11}$ or $\Sigma_{22}$ in the propagator in 
\eqrf{Sig} [cf.\ \eqrf{S}] by the perturbative 
expression ${\Sigma^0}$, \eqrf{self-energy}, the 
difference contributing at sub-subleading order 
to the gap equation. In order to solve the gap equation,  
let us revert the analytic continuation to 
real energies in \eqrf{self-energy}, i.e., 
$q_0$ is purely imaginary in the following. From 
Eqs.\ (\ref{S}) and (\ref{self-energy}), 
the effect of including ${\Sigma^0}$ is to replace 
\bea
q_0\longrightarrow 
\frac{q_0}{Z(q_0)} 
\label{q_0}
\eea
in the quark propagator, where 
\bea
Z(q_0)\equiv 
\left( 1+\bar{g}^2 \ln \frac{M^2}{q_0^2} \right)^{-1}
\label{Z}
\eea
is the quark wave-function renormalization factor. 
Since we only want to 
consider the real part of the quark 
self-energy, we shall ignore the cut of the logarithm in 
\eqrf{Z} when performing the Matsubara 
sum in \eqrf{Sig} by contour integration. 
In other words, we assume that the quark propagator 
has only simple poles in the complex $q_0$-plane, 
corresponding to the excitation energies 
of quasi-particles with infinite lifetime. 
This approximation is valid up to subleading order in the 
gap equation, because, as explained above, 
effects from a finite quasi-particle lifetime enter only 
at sub-subleading order. 

A wave function renormalization of the form 
(\ref{Z}) is known from non-relativistic systems 
\cite{holstein}, where it leads to non-Fermi liquid 
behavior. In relativistic systems, 
non-Fermi liquid behavior has been recently studied 
in great detail by Boyanovsky and de Vega 
\cite{boy00}. 

After these introductory remarks, we may immediately 
proceed to Eq.\ (3.3) of \cite{man1} or 
Eq.\ (32) of \cite{pis00}. This equation 
determines the spin-zero gap 
in a two-flavor color superconductor in pure 
Coulomb gauge. With the replacement (\ref{q_0}) it reads 
\bea
\f (K) &=& \frac 23 g^2 \frac TV \sum_Q 
Z^2(q_0)\; \frac{\f (Q)}{q_0^2-[Z(q_0)\,\e_q]^2}
\bigg[ \D_l(K-Q)
\frac{1+\hat{\bf k}\cdot \hat{\bf q}}{2}\non
&&+\D_t(K-Q)\bigg(
-\frac{3-\hat{\bf k}\cdot \hat{\bf q}}{2}
+\frac{1+\hat{\bf k}\cdot \hat{\bf q}}{2}
\frac{(k-q)^2}{({\bf k}-{\bf q})^2}\bigg) \bigg]\;,
\eea
where we neglected the contribution 
of anti-particles. The next step is to 
perform the Matsubara sum over $q_0$.  
We use spectral representations for the 
propagators, as in \cite{pis00}. The only 
difference to the calculation of \cite{pis00} is 
that the poles of the fermion propagator are 
shifted. To leading order, they are now given by 
\bea
q_0 \simeq \pm Z(\e_q)\;\e_q 
\equiv \pm \tilde{\e}_q\;. 
\eea
The rest of the calculation is straightforward. 
We also take the external quark energy $k_0$ 
to be on the new quasi-particle mass-shell, 
$k_0=Z(\e_k)\e_k=\tilde{\e}_k$. Then, 
in analogy to Eq.\ (3.4) of \cite{man1} and 
Eq.\ (72) of \cite{pis00}, the final result 
for the gap equation,  
including the quark self-energy, reads
\bea
\phi_k & \simeq & \bar{g}^2\, 
\int^{\delta}_{0} \frac{d(q-\m)}{\tilde{\e}_q} 
\; Z^2(\tilde{\e}_q)\; 
\tanh \left( \frac{\tilde{\e}_q}{2T}\right)
\frac 12 \ln \left(\frac{\tilde{b}^2\mu^2} 
{|\tilde{\e}^2_q-\tilde{\e}_k^2|} \right)\; \phi_q \;, 
\label{phik}
\eea 
where $\tilde{b}\equiv 256\p^4[2/(N_f g^2)]^{5/2}$.  
Note that we have replaced the symbol $b$ in Eq.\ (72) of \cite{pis00} 
by $\tilde{b}$, because the definition of $b$, cf.\ Eqs.\
(\ref{b},\ref{b0}), includes $b_0'$, the value of which has 
yet to be determined. 
In Ref.\ \cite{pis00}, this distinction was not necessary, 
because there $b_0' \equiv 1$.
We abbreviated $\f_k\equiv \f (\tilde{\e}_k,{\bf k})$; 
$\f_q$ is defined similarly. 

\section{Solving the gap equation}

Let us now solve the gap 
equation (\ref{phik}) at zero temperature. 
In this case, the factor $\tanh \left[\tilde{\e}_q/(2T)\right] =1$.  
Moreover, to leading order we can make the replacements   
$\tilde{\e}_q\ra \e_q$ and $\tilde{\e}_k\ra \e_k$ 
in the logarithm 
$\ln \left( \tilde{b}^2\mu^2/|\tilde{\e}^2_q-\tilde{\e}_k^2| \right)$.  
For similar reasons, $Z(\tilde{\e}_q)\simeq Z(\e_q)$. 
Following Ref.\ \cite{son99}, we approximate 
\bea
\frac 12 \ln \left( \frac{\tilde{b}^2\mu^2} 
{|\e^2_q-\e_k^2|} \right)\longrightarrow 
\ln \left(\frac{\tilde{b}\m}{\e_q} \right)\;\q (q-k)
+\ln \left(\frac{\tilde{b}\m}{\e_k} \right)\;\q (k-q)\;, 
\eea
and then introduce the variables \cite{pis00}
\begin{mathletters}
\bea
x&=&\bar{g}\;\ln \left( \frac{2\,\tilde{b}\m}{k-\m+\e _k}\right)\;,\\
y&=&\bar{g}\;\ln \left( \frac{2\,\tilde{b}\m}{q-\m+\e _q}\right)\;,\\
x^*&=&\bar{g}\;\ln \left( \frac{2\, \tilde{b}\m}{\f_0}\right) \;,
\label{xyx*}\\
x_0&=&\bar{g}\;\ln \left( \frac{\tilde{b}\m}{\d}\right) \;.
\eea
\end{mathletters}
Note that in contrast to \cite{pis00} we choose 
to include a factor $\bar{g}$ in the definition of these 
variables. Consequently, since $\f_0\sim \m \exp (-1/\bar{g})$, 
$x^*\sim O(1)$ and $x_0\sim O(\bar{g})$. Furthermore, 
$x$ and $y$ are of order $O(1)$ near and of order $O(\bar{g})$ 
away from the Fermi surface. 

In analogy to Eqs.\ (84) and (85) of \cite{pis00}, 
the gap equation and its derivatives 
read in these new variables 
\begin{mathletters}
\bea
\f (x) & \simeq &  
x\int _x^{x^*}\; dy
( 1-2\bar{g}y )\f (y)
+\int _{x_0}^{x}dy\; y 
( 1-2\bar{g}y ) \f (y)\; , 
\label{0-gpeq} \\
\frac{d\f (x)}{dx}&\simeq&
\int _x^{x^*}dy \; ( 1-2\bar{g}y ) \; \f (y)\;,
\label{1st-gpeq}\\
\frac{d^2\f (x)}{dx^2} 
& \simeq &-( 1-2\bar{g}x )\; \f (x)\;.
\label{2nd-gpeq-x} 
\eea 
\end{mathletters}
In these equations, we 
neglected contributions of order 
$O(\bar{g}^2)$, for instance, a term 
$\bar{g}^2\ln (\tilde{b}\m/M)$ 
in the wave function renormalization factor $Z(\e_q)$. 

In order to solve \eqrf{2nd-gpeq-x},  
we replace $x$ with a new variable $z$, 
\be
z \equiv -(2\bar{g})^{-2/3}
\big(\;1-2\bar{g} x \;\big)\;,
\label{zx}
\ee 
and obtain Airy's differential equation \cite{as}, 
\be
\frac{d^2\f (z)}{dz^2}=z\f (z)\;.
\label{2nd-gpeq-z}
\ee
The solution $\f (z)$ of \eqrf{2nd-gpeq-z} 
is a linear combination of the 
Airy functions Ai($z$) and Bi($z$), 
\bea
\f (z)=C_1\;{\rm Ai}(z)+C_2\;{\rm Bi}(z)\;.
\eea
In weak coupling, $z$ is always negative, 
and the Airy functions and their first 
derivatives can be expressed in 
terms of modulus and phase, defined as 
\bea
&&{\rm Ai}(z)=M(|z|)\cos \q (|z|)\; ,\;\;
{\rm Bi}(z)=M(|z|)\sin \q (|z|)\;,\nonumber\\
&&M(|z|)=\sqrt{{\rm Ai}^2(z)+{\rm Bi}^2(z)}\; ,\;\;
\q (|z|)=\arctan\left[ \frac{{\rm Bi}(z)}{{\rm Ai}(z)}\right] 
\; ,\nonumber\\
&&{\rm Ai'}(z)=N(|z|)\cos \vf (|z|)\; ,\;\;
{\rm Bi'}(z)=N(|z|)\sin \vf (|z|)\; ,\nonumber\\
&&N(|z|)=\sqrt{{\rm Ai'}^2(z)+{\rm Bi'}^2(z)}\; ,\;\;
\vf (|z|)=\arctan\left[ \frac{{\rm Bi'}(z)}{{\rm Ai'}(z)}\right]\;. 
\eea
At the Fermi surface, the value of the 
zero-temperature gap function is 
$\f (z^*)=\f_0$ and its derivative vanishes, 
$d\f (z^*)/dz =0$. Consequently, we obtain for 
the gap function 
\be 
\f (z)=\f_0\;\frac{M(|z|)}{M(|z^*|)}\;
\frac{\sin \big[\varphi (|z^*|)-\q (|z|)\big]}
{\sin \big[ \varphi (|z^*|)-\q (|z^*|)\big]}\;.
\label{f-c1}
\ee 

In order to determine $\f_0$, we use \eqrf{0-gpeq} 
at the Fermi surface, $z=z^*$, 
and substitute the integration variable $y$ by 
$u\equiv -(2\bar{g})^{-2/3} (1-2\bar{g} y)$ to obtain 
\be
\f (z^*)=\int _{z^*}^{z_0} du
\big[ u+(2\bar{g})^{-2/3} \big] u \f(u)\;,
\label{fz*}
\ee
where $z_0=-(2\bar{g})^{-2/3}
[1-2\bar{g} x_0 ]$. According to \eqrf{2nd-gpeq-z}, 
we can replace $u\f (u)$ with $d^2\f (u)/du^2$. 
Integrating by parts, this leads to the condition 
\be
\big[z_0+(2\bar{g})^{-2/3}\big]\;\f '(z_0)=\f (z_0)\;. 
\label{f'f}
\ee
Note that the above equation depends on $z^*$ through 
\eqrf{f-c1}. It seems that \eqrf{f'f} 
also depends on $z_0$ which 
is arbitrary and far from the Fermi surface. 
In weak coupling, however, the dependence on 
$z_0$ disappears, as we shall show in the following. 
We first rewrite the condition (\ref{f'f}) as 
\be
\big[z_0+(2\bar{g})^{-2/3}\big]\;
\sin \bigg[ \vf (|z^*|)-\vf (|z_0|) \bigg]
=\frac{M(|z_0|)}{N(|z_0|)}
\sin \bigg[ \vf (|z^*|)-\q (|z_0|) \bigg] \;.
\label{mn1}
\ee
In weak coupling, $|z|\sim (2\bar{g})^{-2/3}\gg 1$, 
and we may use the asymptotic formulas \cite{as}
\bea 
\vf (|z|) & \simeq &
\frac {3\p}{4}-\frac 23|z|^{3/2}
-\frac 7{48}|z|^{-3/2}+O(|z|^{-9/2}) 
\non
& \simeq & -\frac{1}{3\bar{g}}+\frac {3\p}{4}
+x-\bar{g}\bigg(\frac{x^2}{2}+\frac{7}{24}\bigg) +O(\bar{g}^2)\; ,
\nonumber\\ 
\q (|z|)&\simeq&\frac {\p}{4} -\frac 23|z|^{3/2}
+\frac 5{48}|z|^{-3/2}+O(|z|^{-9/2})\nonumber\\
& \simeq & -\frac{1}{3\bar{g}}+\frac{\p}{4}
+x-\bar{g} \bigg(\frac{x^2}{2}-\frac{5}{24}\bigg) +O(\bar{g}^2) 
\; ,\nonumber\\ 
\frac{M(|z|)}{N(|z|)}&\simeq& 
|z|^{-1/2}\bigg[1+O(|z|^{-3})\bigg]
\; , 
\eea 
where we employed $|z|\simeq (2\bar{g})^{-2/3}(1-2\bar{g}x)$, 
cf.\ \eqrf{zx}. We now expand \eqrf{mn1} to order $O(\bar{g})$ 
and obtain 
\be
x^*=\arctan \bigg[-\frac{2}{\bar{g}(1+x^{*2})}\bigg]\;.  
\ee
In weak coupling, the argument of the 
arctan is large, and we can expand 
the right-hand side to order $O(\bar{g})$ 
around $\p /2$. The result is
\be
\label{x*}
x^*\simeq \frac{\p}{2}+\bar{g}\frac{1+x^{*2}}{2} \;.
\ee
To order $O(\bar{g})$, we can 
approximate $x^{*2}\simeq \p ^2/4$ 
on the right-hand side of \eqrf{x*}, 
and using the definition of $x^*$, 
\eqrf{xyx*}, we obtain the zero-temperature 
gap value at the Fermi surface 
\bea
\f _0 = 2\; \tilde{b}\; b_0'\;\m\; 
\exp \bigg(-\frac{\p}{2\bar{g}}\bigg)\;,
\label{34}
\eea
where $b_0'$ is given by \eqrf{b0'}, with $c_1'$ of \eqrf{c1'}. 
In conclusion, the effect of including the
quark self-energy in the gap equation changes 
the value of $b_0'$ from one, as in \cite{SchaferWilczek,pis00}, 
to the value $c_1'$ given in \eqrf{c1'}. 

\section{Determining the transition temperature}

Within the gap equation approach, we can 
also determine the temperature $T_c$ for the transition 
between the normal and the 
superconducting phase. 
We follow Ref.\ \cite{pis00} 
and consider the gap equation 
(\ref{0-gpeq}) at the Fermi surface, 
restoring the factor $\tanh [\e (y)/(2T)]$ present 
at non-zero temperature. 
As in \cite{pis00}, we assume that to leading order 
the shape of the gap function at non-zero temperature 
is the same as at zero temperature, and that only the 
overall magnitude changes with temperature, 
$\f (x,T)=\f (T)\f (x,0)/\f _0$. 
Then the gap equation reads 
\be
\label{gap-t}
1=\int _{x_0}^{x^*}dy\; y (1-2\bar{g}y)  
\tanh \frac{\e (y)}{2T}\;\frac{\f (y,0)}{\f _0}\;.
\ee
We now separate the range of integration into two pieces,  
$[x_0,x^*]\rightarrow [x_0,x_{\k}]+[x_{\k},x^*]$,  
where $x_{\k}=x^*-\bar{g}\ln (2\k)
=\bar{g}\ln [\tilde{b}\m/(\k\f _0)]$. 
The main contribution to the integral in 
\eqrf{gap-t} comes from the region of momenta
away from the Fermi surface, $[x_0,x_{\k}]$.
In this region, $\e (y)\gg \f_0\sim T$, 
such that we can approximate  
the factor $\tanh [\e (y)/(2T)]\simeq 1$.  
By making use of \eqrf{2nd-gpeq-z} and \eqrf{f'f},  
the integral over the region $[x_0,x_{\k}]$ is evaluated as 
\bea
{\cal I}&=&\int _{x_0}^{x_{\k}}dy\; y
( 1-2\bar{g}y ) \;\frac{\f (y,0)}{\f _0} 
=\frac{1}{\f _0}
\bigg\{ \f (z_{\k})-\Big[ z_{\k}+(2\bar{g} )^{-2/3}\Big]
\f '(z_{\k})\bigg\}
\non
&=&1-\frac{\p}{2}\bar{g}\ln 2\k+O(\bar{g}^2) \;,
\eea
where $z_{\k}=z^*-(2\bar{g})^{1/3}\bar{g}\ln 2\k$. The last line 
is obtained by expanding the right-hand side 
of the second equality to order $O(\bar{g})$ around $z^*$. 
Equation (\ref{gap-t}) becomes 
\bea
\int _{x_{\k}}^{x^*}dy\; y (1-2\bar{g}y)  
\tanh \frac{\e (y)}{2T}\;\frac{\f (y,0)}{\f _0}
=\frac{\p}{2}\bar{g}\ln 2\k+O(\bar{g}^2)\;.
\eea
The integral on the left-hand side may 
now be computed to order 
$O(\bar{g})$. As in \cite{pis00}, 
this amounts to approximating 
$y\simeq x^*$ and $\f (y,0)/\f_0\simeq 1$. Furthermore,  
the correction from the quark self-energy can be neglected, 
$1-2\bar{g}y\simeq 1$. 
In this way, we obtain Eq.\ (104) of 
\cite{pis00}; consequently the BCS result (\ref{Tcphi}) 
remains valid to leading order in $g$, even when 
the quark self-energy is taken into account in the gap 
equation. With Eqs.\ (\ref{Tcphi}) and 
(\ref{34}), we thus conclude that our result for $T_c$ 
is the same as that obtained in Ref.\ \cite{bro00}.

\section{Conclusions}

In this paper, we have computed the 
spin-zero gap in a two-flavor color superconductor at 
zero temperature from a mean-field gap equation. 
In contrast to earlier studies 
\cite{SchaferWilczek,pis00,hong,miransky,man1}, we 
have included subleading contributions from the 
real part of the quark self-energy.  
We found that these contributions reduce 
the gap parameter at the Fermi surface by a factor 
$b_0'=\exp [-(\p ^2+4)/8]\simeq 0.177$. 
We then computed the transition temperature 
$T_c$ between the normal and superconducting 
phase and found that the BCS relation 
$T_c\simeq 0.57\f_0$ remains valid to leading order in 
$g$ after including the corrections from 
the quark self-energy. Therefore, we obtain 
the same value for $T_c$ as in Ref.\ \cite{bro00}.

\acknowledgments

The authors thank R.\ Pisarski and 
I.\ Shovkovy for discussions and 
a critical reading of the manuscript. 
Q.\ W.\ acknowledges financial support from 
Alexander von Humboldt-Foundation. 
He also appreciates help and support from 
the Institut f\"ur Theoretische Physik of the 
J.\ W.\ Goethe-Universit\"at, and especially from 
Prof.\ W.\ Greiner and Prof.\ H.\ St\"ocker.


\begin{thebibliography}{99}

\bibitem{asymp}
D.J.\ Gross and F.\ Wilczek, Phys.\ Rev.\ Lett.\ {\bf 30}, 
1343 (1973); H.D.\ Politzer, ibid.\ 1346. 

\bibitem{bai84}
D.\ Bailin and A.\ Love, 
Phys.\ Rep.\ {\bf 107}, 325 (1984).

\bibitem{BCS} 
J.R.\ Schrieffer, {\it Theory of Superconductivity}
(New York, W.A.\ Benjamin, 1964).

\bibitem{SchaferWilczek}
T.\ Sch\"afer and F.\ Wilczek, 
Phys.\ Rev.\ D {\bf 60}, 114033 (1999).  

\bibitem{pis00}
R.D.\ Pisarski and D.H.\ Rischke,
Phys.\ Rev.\ D {\bf 61}, 
051501, 074017 (2000).  

\bibitem{ris01} D.H.\ Rischke, 
Phys.\ Rev.\ D {\bf 64}, 094003 (2001). 

\bibitem{son99}
D.T.\ Son, Phys.\ Rev.\ D {\bf 59}, 094019 (1999).

\bibitem{hong}
D.K.\ Hong, Phys.\ Lett.\ B {\bf 473}, 118 (2000); 
Nucl.\ Phys.\ {\bf B 582}, 451 (2000).  

\bibitem{miransky}
I.A.\ Shovkovy and L.C.R.\ Wijewardhana, 
Phys.\ Lett.\ B {\bf 470}, 189 (1999); 
D.K.\ Hong, V.A.\ Miransky, I.A.\ Shovkovy, 
and L.C.R.\ Wijewardhana, Phys.\ Rev.\ D {\bf 61}, 056001 (2000); 
Erratum, ibid.\ D {\bf 62}, 059903 (2000).  



\bibitem{ShusterRajagopal}
K.\ Rajagopal and E.\ Shuster, Phys.\ Rev.\ D 
{\bf 62}, 085007 (2000).  

\bibitem{man1} C.\ Manuel,
Phys.\ Rev.\ D {\bf 62}, 114008 (2000). 

\bibitem{bro00} W.E.\ Brown, J.T.\ Liu, and  
H.-C.\ Ren, Phys.\ Rev.\ 
D {\bf 61}, 114012 (2000); ibid.\  
D {\bf 62}, 054013, 054016 (2000).  

\bibitem{pis99-2}
R.D.\ Pisarski and D.H.\ Rischke,
Phys.\ Rev.\ D {\bf 60}, 094013 (1999).  

\bibitem{man0} C.\ Manuel, 
Phys.\ Rev.\ D {\bf 62}, 076009 (2000).  


\bibitem{holstein}
T.\ Holstein, R.E.\ Norton, and P.\ Pincus, 
Phys.\ Rev.\ B {\bf 6}, 2649 (1973).


\bibitem{boy00} D.\ Boyanovsky and H.J.\ de Vega, 
Phys.\ Rev.\ D {\bf 63}, 034016 (2001).  

\bibitem{as} {\it Handbook of Mathematical Functions}, eds.  
M.\ Abramowitz, I.A.\ Stegun (Dover, New York, 1965). 


\end{thebibliography}
\end{document}